\documentclass[runningheads]{llncs}
\usepackage[T1]{fontenc}

\usepackage{graphicx}
\usepackage{subfigure}
\usepackage{amsmath}
\usepackage{multirow}
\usepackage{array}
\usepackage{bbm}
\usepackage{xcolor}
\usepackage[linesnumbered,ruled,vlined]{algorithm2e}
\usepackage[noend]{algpseudocode}

\SetCommentSty{mycommfont}

\begin{document}

\title{Query Generation based on Generative Adversarial Networks}

 \author{Weihua Sun \and
 Run-An Wang \and
 Zhaonian Zou}

 \authorrunning{W. Sun et al.}

 \institute{Harbin Institute of Technology, China}

\maketitle

\begin{abstract}
Many problems in database systems, such as cardinality estimation, database testing and optimizer tuning, require a large query load as data. However, it is often difficult to obtain a large number of real queries from users due to user privacy restrictions or low frequency of database access. Query generation is one of the approaches to solve this problem. Existing query generation methods, such as random generation and template-based generation, do not consider the relationship between the generated queries and existing queries, or even generate semantically incorrect queries. 
In this paper, we propose a query generation framework based on generative adversarial networks (GAN) to generate query load that is similar to the given query load. In our framework, we use a syntax parser to transform the query into a parse tree and traverse the tree to obtain the sequence of production rules corresponding to the query. The generator of GAN takes a fixed distribution prior as input and outputs the query sequence, and the discriminator takes the real query and the fake query generated by the generator as input and outputs a gradient to guide the generator learning. In addition, we add context-free grammar and semantic rules to the generation process, which ensures that the generated queries are syntactically and semantically correct. We conduct experiments to evaluate our approach on real-world dataset, which show that our approach can generate new query loads with a similar distribution to a given query load, and that the generated queries are syntactically correct with no semantic errors. The generated query loads are used in downstream task, and the results show a significant improvement in the models trained with the expanded query loads using our approach.
\end{abstract}

\section{Introduction}
\label{sec:introduction}

In a database system, a query load refers to a set of queries with a specific distribution. Many problems in database systems, such as cardinality estimation~\cite{dutt2019selectivity}, database testing~\cite{zhong2020squirrel}, view/index selection~\cite{yuan2020automatic}, database partitioning~\cite{curino2010schism}, and optimizer tuning~\cite{li2019qtune}, require a large query load to participate in method building or testing. For example, as a fundamental problem in query optimization, cardinality estimation predicts how many tuples will appear in the result of a query, i.e. the cardinality, before the query is actually executed. A learning-based query-driven cardinality estimator uses a machine learning model to map a query to its cardinality. Such an estimator requires training the model with a large number of high-quality queries as well as their cardinalities. If there are no sufficient queries in the training set, the model's generalization performance of tends to be poor. Moreover, when new query loads arrive, the estimator needs to be updated to keep the model accurate and up-to-date.

Query loads play very important roles in solving various problems in database systems. However, except large IT enterprises, it is actually difficult to obtain a large number of real queries from users due to user privacy or low database access frequency. Query generation, the technique of generating queries for a database, is one way to address this problem. Query generation can either be independent of any given query loads or depend on a known query load. The existing query generation methods, such as random generation~\cite{slutz1998massive,sqlsmith} and template-based generation~\cite{bruno2006generating,slutz1998massive}, do not consider the relationships between the generated query load and the given query load, and even the generated queries may have errors in syntax or semantics. Therefore, it is really necessary to design a query-load-dependent method to generate a query load with a very similar distribution to a given query load, so as to expand the given query load.

Query generation bears some similarity to image or text generation in the field of artificial intelligence. With the development of machine learning, generative models using deep learning are being studied by a growing number of works. Generative models such as generative adversarial networks (GAN)~\cite{goodfellow2020generative} and variational autoencoder (VAE)~\cite{kingma2013auto}, have achieved good results in image generation~\cite{karras2019style} and text generation~\cite{yu2017seqgan}. In some text generation tasks with grammar, TreeGAN~\cite{liu2018treegan} and other works~\cite{yin2017syntactic} have put forward preliminary attempts. However, these methods have not been studied for specific languages such as SQL, and have not added domain knowledge to the generation process.

In this paper, we propose a GAN-based query generation framework to generate a query load that is similar to the given query load. In our framework, we use a syntax parser to transform the query into a parse tree and traverse the tree to obtain the sequence of production rules corresponding to the query. The generator of GAN takes a fixed distribution prior as input and outputs the query sequence, and the discriminator takes the real query and the fake query generated by the generator as input and outputs a gradient to guide the generator learning. We also define a simplified version of the SQL context-free grammar (CFG) to guide query generation. A matrix of masks based on SQL CFG ensures that the generated queries are syntactically correct. Moreover, we summarize common semantic errors and implement semantic-rule-based generative constraints to avoid generating semantically wrong queries. During generation, we keep track of the semantic state of the current sequence and mask actions that will cause semantic errors. To enable the model to better learn the query distribution, we add query feature information, such as query cardinality and estimated cost, to the query sequence. Correspondingly, we modify the discriminator of the GAN to capture this information. 

The contributions of this paper can be summarized as follows:
\begin{enumerate}
    \item We propose a GAN-based query generation method that can learn the distribution of a given query load and generate new query loads.
    \item We define a simple SQL CFG and apply it to the query generation process so that the generated queries syntax are correct. We summarize common semantic errors and implement semantic-rule-based generative constraints to ensure the semantic correctness of the generated queries.
    \item We encode the feature information of the query and add it to the generation process to improve the quality of the generated query.
    \item Experiments show that our model can generate high-quality queries and the generated query load is similar to the given query load distribution. We apply the generated query load to cardinality estimation task, and results show a significant improvement in model effectiveness using the generated query expansions' query loads as training data.
\end{enumerate}


\section{Problem Statement}
\label{sec:problem-statement}

Let $D = \{R_1, R_2, \dots, R_t\}$ be a relational database, where $R_1, R_2, \dots, R_t$ are relations. Given an input query load $Q = \{q_1, q_2, \dots, q_n\}$ on $D$ and $m \in N$, the query-load-dependent query generation problem generates another query load $Q' = \{q'_1, q'_2, \dots, q'_m\}$ such that $Q'$ is very similar to $Q$. The formulation of the similarity between $Q$ and $Q'$ is essential for query generation, which is defined as follows:

A query $q$ is an SQL statement. There are a lot of features to characterize a query $q$. Simple features include the number of tables, joins, predicates, and attributes involved in $q$; Complex features include the cardinality of $q$, the cost of $q$ estimated by the query optimizer of the DBMS, and so on. Let $f(q)$ be the value of a feature $f$ of a query $q$. For a query load $X$, the distribution of $f(q)$ for all queries $q \in X$ characterizes a specific aspect of $X$ in terms of the feature $f$. Let $p^f_X$ denote the probability density function (pdf) of this distribution. Therefore, we can evaluate the distance between the input query load $Q$ and the generated query load $Q'$ in terms of the feature $f$ by the divergence between $p^f_{Q}$ and $p^f_{Q'}$, that is,
\begin{equation}\label{eqn:dist-f}
	d^f(Q, Q') = \int_x \left|p^f_{Q}(x) - p^f_{Q'}(x)\right| dx,
\end{equation}
where $x$ is taken from the set of all possible values of $f(q)$ for all $q \in X$.

Let $F = \{f_1, f_2, \dots, f_k\}$ be the set of features for characterizing queries. The joint distribution of $f_1(q), f_2(q), \dots, f_k(q)$ for all queries $q \in X$ is a characterization of $X$. We use $p^F_X$ to denote the pdf of this joint distribution. In general, the distance between the input query load $Q$ and the generated query load $Q'$ can be evaluated by
\begin{multline}\label{eqn:dist-all}
	d^F(Q, Q') = \\
	\int_{x_k} \int_{x_{k - 1}} \dots \int_{x_1} \left|p^F_{Q}(x_1, x_2, \dots, x_k) - p^F_{Q'}(x_1, x_2, \dots, x_k)\right| dx_1 d_{x_2} \dots dx_k,
\end{multline}
where $x_i$ is taken from the set of all possible values of $f_i(q)$ for all $q \in X$.

Therefore, the problem to be solved in this paper can be formally stated as follows: Given a query load $Q = \{q_1, q_2, \dots, q_n\}$ on $D$, $m \in N$, and a set $F$ of query features, we expect to generate a query load $Q' = \{q'_1, q'_2, \dots, q'_m\}$ that minimizes the distance $s^F(Q, Q')$ between $Q$ and $Q'$. In addition, the generated queries $q'_1, q'_2, \dots, q'_m$ are required to be syntactically and semantically correct.

\section{Our Method}
\label{sec:our-method}

\begin{figure}[htp]
	\centering
	\includegraphics[width=0.99\textwidth]{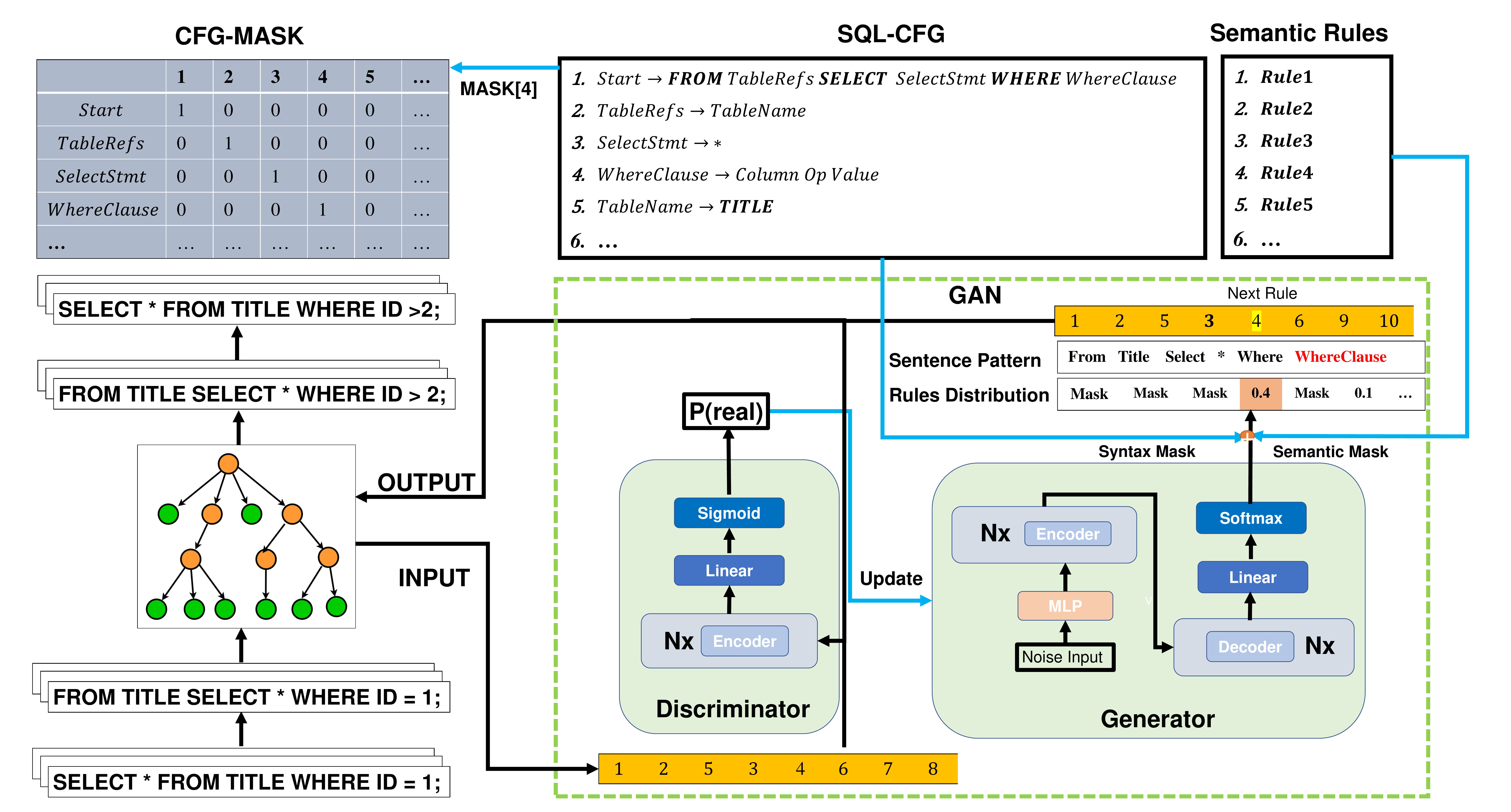}
	\caption{The Model Architecture.}
	\label{fig:architecture}
\end{figure}

In this paper, we propose a query generation method based on GAN. The architecture of our model is shown in Fig.~\ref{fig:architecture}. First, we preprocess the given SQL queries in $Q$ (Section~\ref{sub:preprocessing}), convert each SQL query into a parse tree, and traverse the  parse tree to obtain its corresponding production rule sequence (Section~\ref{sub:grammar-guidance}). Next, we feed the production sequences into the model as training data, and the fully trained generator can capture the distributional information of the queries and generate queries similar to the given query load $Q$ (Section~\ref{sec:generative-model}). During the training and inference stage, syntax and semantic rules are used to constrain the generator to generate correct queries (Section~\ref{sec:semantic-constraints}). Finally, the model converts the generated production sequences to SQL queries.

\subsection{Preprocessing}
\label{sub:preprocessing}
In our method, we first carry out preprocessing on the queries in the given query load $Q$. Preprocessing include the following procedures:
\paragraph{Restructure the query}
We adapt the query structure to make it easier to design syntax and semantic rules. The query structure supported by the generator is \texttt{SelectField FromField WhereClause GroupClause HavingClause}, we adapt it to \texttt{FromField SelectField WhereClause HavingClause} \texttt{GroupClause}. We make two adjustments, first we move the \texttt{FromField} forward to the beginning of the query, because the \texttt{FromField} qualifies relations involved in the query. Secondly, we move the \texttt{GroupClause} to the end because the attributes that appear in \texttt{SelectField} and \texttt{HavingClause} but are not aggregated need to be grouped in the \texttt{GroupClause}.

\paragraph{Reduce action space}
There are often many constant values in a query. These constants make the action space of query generation extremely large. To reduce the action space, we construct a histogram for each attribute involved in the queries in $Q$, that is, we divide the domain of the attribute into buckets and represent each bucket with a distinct hash key. Then, we replace each constant in a query with the hash key of the bucket that the constant belongs to.

After preprocessing the query as described above, the semantics of the query does not change. In next stpdf of our method, queries are also generated in this form. Reducing a preprocessed query to executable SQL requires only two stpdf: the first step reverts \texttt{FromField} and \texttt{GroupClause}, and the second step randomly samples a value from the hash bucket based on the hash key and replaces the hash key with this value.

\subsection{Grammar Guidance}
\label{sub:grammar-guidance}
The common text generation work generates sequences directly, without considering the structure and grammar of the sequence. If the generator can generate sequences according to the grammar, we can get syntactically correct text sequences. In this section, we will describe how to convert a query into a tree structure and how to generate queries based on grammars.

The grammar of SQL can be represented by context-free grammar (CFG). We use the CFG of SQL to guide query generation. A CFG is defined as
\begin{equation*}
	CFG = \{V, T, P, S\}
\end{equation*}
where $V$ is the set of non-terminal symbols, $T$ is the set of terminal symbols, $P$ is the set of productions of the form $P \rightarrow \gamma$, where $P \in  V, \gamma \in (V \cup T)^*$, and $S$ is the start symbol of the grammar. In our work, we make a basic CFG for SQL. A simplified version of this grammar is shown in Table~\ref{tab:cfg}.
\begin{table}[h!]
	\caption{Simplified SQL CFG}
	\label{tab:cfg}
	\centering
	\begin{tabular}{l} 
		\hline
		0:  $Start \rightarrow \boldsymbol{FROM}\ TableRefs\ \boldsymbol{SELECT}\ SelectStmt\ \boldsymbol{WHERE}\ WhereClause$ \\
            1:	$TableRefs\rightarrow TableName$\\
            2:  $SelectStmt\rightarrow  \boldsymbol{\ast}$\\
            3:  $WhereClause\rightarrow Column\ Op\ Value$\\
            4:  $TableName\rightarrow \boldsymbol{TITLE}$\\
            5:  $Column\rightarrow\boldsymbol{ID}$\\
            6: $Op\rightarrow\boldsymbol{=}$\\
            7: $Value\rightarrow\boldsymbol{1}$\\
            8:  $\cdots$\\
		\hline
	\end{tabular}
\end{table}

\begin{figure}[htp]
    \centering
    \includegraphics[width=0.7\textwidth]{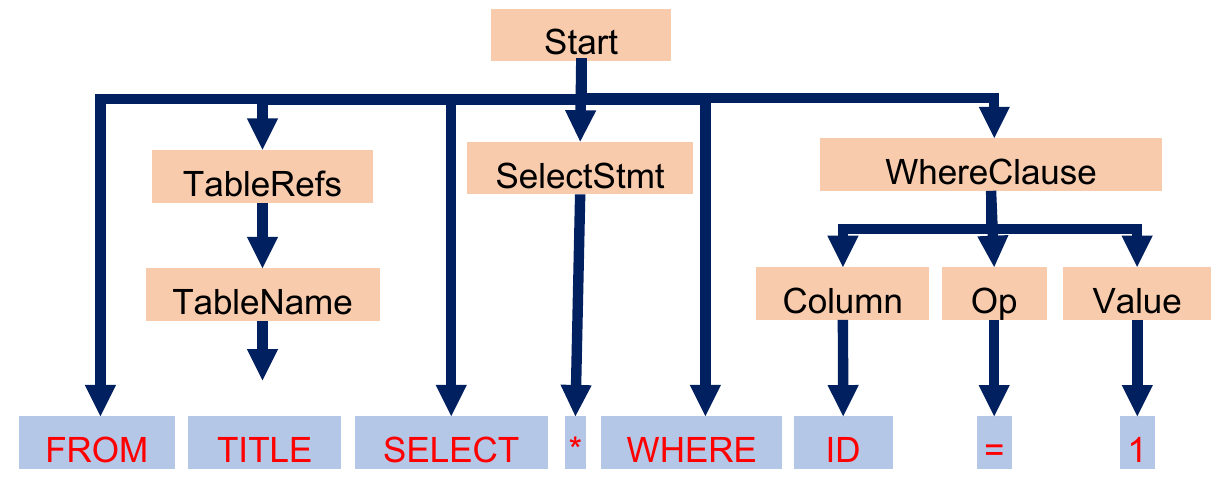}
    \caption{Parser Tree for "FROM TITLE SELECT * WHERE ID = 1"}
    \label{fig:parser tree}
    \
\end{figure}

Algorithm~\ref{al:1} shows how to convert the query load to production sequences. For each query, the model uses a parser to convert it into a parse tree, and then performs a depth-first, left-to-right traversal of the tree. During the traversal: if a non-terminal symbol is encountered, the non-terminal is used as the head of the production and its children as the body of the production, then the combined production is added to the production sequences. Finally, it outputs the production sequences corresponding to the input query load. Fig.~\ref{fig:parser tree} shows the parse tree of query "FROM TITLE SELECT * WHERE ID = 1", and by traversing the parse tree we can get the generated sequence [0,1,4,2,3,5,6,7].

\begin{algorithm}[ht]
\label{al:1}
  \DontPrintSemicolon
  \SetKwInput{KwInput}{Input}                
  \SetKwInput{KwOutput}{Output}              
  \SetKwFunction{FMain}{Main} 
  \SetKwFunction{FDepthFirstTraversal}{DepthFirstTraversal} 
  \SetKwFunction{FCheckNode}{CheckNode} 
  \SetKwFunction{FParser}{Parser} 
  \KwInput{Query load $Q = \{q_1, q_2, \dots, q_n\}$, SQL-CFG $G = \{V, T, P, S\}$}
  \KwOutput{Production Sequences  $S = \{s_1, s_2, \dots, s_n\}$}
 \SetKwProg{Fn}{Function}{:}{\KwRet}
 \Fn\FMain{
  \For{$q_i$ in $Q$}
  {
    tree = \FuncSty{Parser($G$,$q_i$)} \tcp{use the LALR parser}
    $s_i$ = \FuncSty{\FDepthFirstTraversal{tree}}\;
    $S = S \cup \{s_i\}$\;
  }
  \KwRet  $S$
 }
 \Fn{\FDepthFirstTraversal{node}}{
      $s = \{ \}$\;
      \FuncSty{\FCheckNode{node,s}}\;
      \For{child in node.children}
      {
        \FuncSty{\FCheckNode{child.root,s}}\;
      }
      
      \KwRet  $s$\;
 }

 \Fn{\FCheckNode{node,s}}{
      \If{node.type is non-terminal}
        {
            head = node \;
            body = node.children.data\;
            $s = s \cup \{head\rightarrow body\}$\;
        }
      \KwRet  $s$
 }
\caption{From queries to production sequences}
\end{algorithm}

Given a CFG $\{V, T, P, S\}$, the derivation process is to convert a sentence pattern of the form $\alpha P \beta$ to $\alpha \gamma \beta$ if and only if there is a production $P \rightarrow \gamma$. Intuitively, CFG starts from the start symbol, and gradually applies the production to the derivation process: find the production whose head matches the leftmost non-terminal in the current sentence pattern, and use the body of the production to replace the matching non-terminal until there is no non-terminal in the sentence pattern. Finally, we get a string that follows this CFG.

Algorithm~\ref{al:2} implements the process from production sequence to query. For each production sequence, the model maintains a symbol stack and a text sequence. The symbol stack stores the sentence pattern to be derived, and initially the stack contains only the start symbol of the grammar. The symbols in the stack are popped from the stack one by one, and if the popped symbol is a non-terminal, the production in the sequence is applied to complete the derivation process once. If a terminal is encountered, terminal is added to the text sequence. If there are no symbols in the symbol stack, the text sequence is the query we need.

An important idea of our query generation method is to exploit the production sequence for deriving a query. For each query $q_i \in Q$, we obtain the production sequence $S_i$ using Algorithm~\ref{al:1}. The production sequence $S_i$ partially reveals the grammatical structure of $q_i$. Our query generation model aims at generating queries with production sequences that are similar to the production sequences $S_1, S_2, \dots, S_n$ for the query load $Q$.
\begin{algorithm}[htp]
\label{al:2}
  \DontPrintSemicolon
  \SetKwInput{KwInput}{Input}                
  \SetKwInput{KwOutput}{Output}              

  \KwInput{Production Sequences  $S = \{s_1, s_2, \dots, s_n\}$}
  \KwOutput{Query load $Q = \{q_1, q_2, \dots, q_n\}$}

 \SetKwProg{Fn}{Function}{:}{\KwRet}
 
    $Q = \{ \}$\;
  \For{$s_i$ in $S$}
  {
    $q_i = \{ \}$\;
    
    Stack symbols =  [Start]\tcp{Derivation from the start symbol of CFG}
    Stack productions = reversed($s_i$)\tcp{From the first production}
   \While{symbols is not empty}
   {
      symbol = symbols.pop()\;
      \If{symbol is terminal}
      {
         $q_i = q_i \cup \{symbol\}$\;
         continue\;
      }
      \Else
      {
         production = productions.pop()\;
        \For{symbol in reversed(production.body)}{
            symbols.push(symbol)\;
      }
      
   }
    
  }
  $Q = Q \cup \{q_i\}$\;
  }
  \KwRet $Q$
\caption{From production sequences to queries}
\end{algorithm}

\subsection{Generative Model}
\label{sec:generative-model}

At the heart of our query generation framework is the generative model represented as a generative adversarial network (GAN). GAN is an unsupervised machine learning framework consisting of a generator network (G) and a discriminator network (D). The generator G takes a fixed prior $z$ as input and outputs a production sequence which can be converted into an SQL query. The discriminator D is a binary classification network that takes a production sequence as input and outputs the probability that the input sequence comes from the input query load rather than G. We want the sequence generated by G to be as similar as possible to the sequence from the given query load, which requires a well-performing D to guide the work of G. Thus, the optimization objective of GAN is:
\begin{equation}
\label{eq:gan}
\mathop{\min}\limits_{G} \mathop{\max}\limits_{D}V\left( D,G \right) =\mathop E \limits_{x \sim p_{data}\left( x \right)}\left[ \log D\left( x \right) \right] +\mathop E \limits_{z \sim p\left( z \right)}\left[ \log \left( 1-D\left( G\left( z \right) \right) \right) \right],
\end{equation}
where $x$ is a production sequence taken from the production sequences of the queries in $Q$. For the optimal D, Eq.~\eqref{eq:gan} can minimize the JS divergence between the generated sequence and the given sequence. To improve the quality of the generated sequences, we train GAN using the method proposed in SeqGAN~\cite{yu2017seqgan}. In the remainder of this section, we discuss the details of the generator and the discriminator.

\subsubsection{Generator}
Inspired by the success of attention mechanism in NLP~\cite{NIPS2017_3f5ee243}, we use a transformer as the generator to generate production sequences which can be converted into queries. The encoder of the transformer takes a fixed prior $z \sim p(z)$ as input, encode it and pass it to decoder. At each step $t$, the input of decoder is 
$\mathbf{x}_t =[\mathbf{p}_{1},\mathbf{p}_{2},\cdots,\mathbf{p}_{t-1}]$, 
where $\mathbf{p}_{i}$ and $i<t$
is the embedding vector of the production chosen at step $i$. The output of the decoder is ${o_t}$, and the network maps $o_t$ to a probability vector $V^t_{1:|P|}$ using a softmax output layer:
\begin{equation}
\label{eq:pvec0}
    V^t_{1:|P|} = softmax({W_vo_t}+{b_v}).
\end{equation}

At each time step $t$, the model samples a production $\mathbf{p}_t$ from the production set $P$ based on vector $V^t_{1:|P|}$. During this process, we apply the chosen productions for derivation until all non-terminals have been replaced with terminals, and we get a production sequence that can be converted into an query.

\subsubsection{Discriminator}
As with the generator, we use a transformer encoder as the discriminator. The encoder takes a production sequence as input and produces an output sequence. We perform an average pooling operation on the output sequence to obtain an embedding representation of the input sequence $o_T$. Then we apply the Sigmod activation function on it to get the probability $\hat{y}$ that the sequence is from the input query load:
\begin{equation*}
    \hat{y} = \sigma(W_o \cdot o_T + b_o).
\end{equation*}

To make the model better learn the distribution of the query, the feature information, such as query cardinality and cost, is also input into the discriminator. In practice, we reconstruct the SQL based on the production sequence, and then input the SQL statement into the database for execution to obtain the feature information. The feature information is simply normalized and encoded to the beginning of the production sequence. Therefore, the input of the discriminator becomes $\mathbf{x} =[\mathbf{f}_{1},\cdots,\mathbf{f}_{n},\mathbf{p}_{1},\cdots,\mathbf{p}_{T}]$, where $\mathbf{f}_{i}$ is the normalized feature information.

\subsection{Syntax and Semantic constraints}
\label{sec:semantic-constraints}

Note that during derivation, we always choose the leftmost non-terminal $v$ in the current sentence pattern and look for the production whose head is also $v$. In the previous section we sample production in the whole production space according to  $V^t_{1:|P|}$ in eq.~\ref{eq:pvec0}. However, the model may select a production whose head is not $v$ and cause errors in the derivation process. For example, the current sentence pattern is "FROM TITLE SELECT * \textbf{WhereClause}" and the leftmost non-terminal symbol is \textbf{WhereClause}, but the generator select the production $TableName\rightarrow Title$, leading to the failure of the derivation process.

To avoid this, we build a CFG-based mask matrix $M\in\{0,1\}^{|V|\times|P|}$, where $|V|$ and $|P|$ are the numbers of non-terminals and productions, respectively. Specifically, for the a non-terminal $v\in V$, its mask vector is denoted as $M(v)$, in which productions with $v$ as head are marked with 1, and other productions are marked with 0. For example, the mask vector of the CFG start symbol $M(\boldsymbol{Start})$ is $(1, 0, 0, \dots, 0)$ because only the 0-th production in Table~\ref{tab:cfg} has $\boldsymbol{Start}$ as its head. At step $t$, the modified probability vector becomes
\begin{equation}
\label{eq:pvec1}
    \dot{V}_{1:|P|}^t = V^t_{1:|P|} \odot M(v),
\end{equation}
where $v$ is the leftmost nonterminal in the current sentence, $\odot$ denotes the element-wise multiplication, and $M(v)$ masks invalid productions.

Although the CFG-based mask matrix can ensure the correct syntax of queries, semantic errors may occur. For example, we may generate a query like "SELECT Name FROM Title WHERE Name = A", and although the query appears to have no syntax errors, if the Name attribute does not exist in the Title relation, the query will not execute correctly. In our work, we implement semantic-rule-based generative constraints to ensure the semantic correctness.

During generation, we keep track of the semantic state of the current sequence and mask productions that will cause semantic errors. At step 0, we initialize a semantic mask $m_{s}=\{1\}_{1:|P|}$. Whenever the generator chooses a production, we update $m_{s}$ based on the semantic rules and the current sequence. When the generator is going to choose a production, its probability vector is modified as:
\begin{equation}
\label{eq:pvec2}
\ddot{V}_{1:|P|}^t = \dot{V}^t_{1:|P|} \odot m_s
\end{equation}
where $\dot{V}^t_{1:|P|}$ is the probability vector given in Eq.~\eqref{eq:pvec1}. In this vector, the productions that cause syntactic and semantic errors are marked with 0, and the generator ignores these productions.

\begin{example}
When we generate a partial query "FROM Title", we update the semantic mask so that the attributes in Title relation are marked as 1 and other attributes are marked as 0. When choose attributes in the relations during the generation of the remaining query, we choose only those attributes that are present in Title relation. This ensures that the query semantics are correct.
\end{example}

\section{Evaluation}
\label{sec:evaluation}
In this section, we perform experiments to evaluate the similarity between the generated query load and the given query load. We also apply the generated queries to the cardinality estimation task to evaluate the availability of the queries.
\subsection{Experiment Setting}
\label{sub:experiment-setting}
\paragraph{Datasets}
We conducted our evaluation on the real-world dataset IMDB. We created the IMDB database and used four randomly initialized generators to synthesize 4 query loads. Each load contains 2000 queries and has a different level of complexity. The information of the 4 query loads is shown in Table~\ref{tab:queryload}.

\begin{table}[htp]
    \centering
    \begin{tabular}{|c|c|c|c|c|c|c|}
    \hline
         Query load &  Nested & Aggregate & String & Query Length & Database schema&Size\\ \hline
         SPJ-N&  No & No&No&[0,40]& 6 Tables and 9 Columns&2000\\\hline
         AN-N& Yes & Yes&No&[0,40]& 6 Tables and 9 Columns&2000\\\hline
         AN-S&Yes & Yes&Yes&[0,40]& 21 Tables and 22 Columns&2000\\\hline
         AN-SL &Yes & Yes&Yes&[40,100]& 21 Tables and 22 Columns&2000\\\hline
    \end{tabular}
    \caption{Query load information}
    \label{tab:queryload}
\end{table}
\paragraph{Baselines}
\begin{enumerate}
        \item Our GAN-based query generation framework, donate as GAN.
        \item Our GAN-based query generation framework, but without encoding query feature information, donate as as GAN-.
	\item The random method generates SQL queries by randomly selecting productions under our defined CFG, and uses semantic rules to ensure that the generated queries are syntactically correct.
	\item The template-based method uses the templates we extracted from the given query load and randomly generates the constants in the predicates. 
\end{enumerate}

For each the query load, we use the baselines to generate 2000 queries. So we have four synthetic query loads and one real query load for each real query load.

\paragraph{Configurations}

The action space of the query generation problem is limited by the grammar and therefore not very large. So the generator and the discriminator use two lightweight transformers with hyper N=6, H=4 and $d_{model}$=128. According to SeqGAN~\cite{yu2017seqgan}, we pre-train the generator and discriminator using the maximum likelihood method. To avoid generating many queries that are identical to some input queries, we use a dropout of 0.3 and $L_2$ regularization with $\lambda = 0.1$ to avoid model overfitting. We also use gradient penalty to avoid pattern collapse while making the training of GAN easier to converge. The learning rates of the generator and the discriminator are 0.001 and 0.0001, respectively.

\subsection{Evaluation of Generated Query Loads}

We first evaluate the quality of the generated query loads. To do this, we compare the input query load $Q$ with the generated query load $Q'$ in terms of many features including character-level similarity, the occurrences of an aggregate functions, the occurrences of an operator, the occurrences of nested query, the number of joined tables, the length of the query, the cardinality of the query, the cost of the query estimated by the query optimizer of the DBMS, and the correlation between attributes.

For a specific feature $f$, let $f(q)$ be the value of the feature $f$ of a query $q$. In our evaluation, we compare the difference between the distribution of $f(q)$ for all queries $q \in Q$ and the distribution of $f(q')$ for all queries $q' \in Q'$. We use two metrics in our evaluation:
\begin{enumerate}
	\item Wasserstein Distances(WD): WD are metrics on probability distributions inspired by the problem of optimal mass transportation. 
    For two cumulative probability functions (cdf) $g_1$ and $g_2$ on the same domain, the WD is  $$ W (g_1,g_2) =  \int_{\mathbbm{R}} |g_1-g_2|dx$$
    where $x$ is taken from the domain of $g_1$ and $g_2$. The larger the WD, the greater the difference between the two distributions.
	\item Maximum Mean Discrepancy(MMD): MMD is also often used to measure the distance between two distributions and can be used for high-dimensional distributions. For two distribution $Q$ and $Q^{'})$ on the same domain, the MMD is denoted as:
    $$MMD(Q,Q^{'},F) = ||\frac{1}{n}\sum_{i=1}^n{F(q_i)}-\sum_{j=1}^m{F(q^{'}_j)}||_{\mathcal{H}}^2$$ where F is the kernel function that maps the input distribution to the Hilbert space. The larger the MMD, the greater the difference between the two distributions.

\end{enumerate}

\begin{table}[htp]
    \centering
    \begin{tabular}{|c|c|c|c|c|c|} \hline
         Query Loads &Methods&Sequence MMD& Cardinality WD&Cost WD &Length WD \\ \hline
         \multirow{4}*{SPJ-N}
         &  GAN         &\textbf{0.0077}     &\textbf{0.3665}     &\textbf{0.3463}     & 0.843     \\\cline{2-6}
         &  GAN-        &0.0098     &0.6665     &0.3573     & 0.945     \\\cline{2-6}
         &  Random      &0.2895     &2.8777     &2.1119     & 5.138     \\\cline{2-6}
         &  Template    &0.0195     &0.6736     &0.4673     & \textbf{0}        \\  \hline
         \multirow{4}*{AN-N}
         &  GAN         &\textbf{0.0129}     &\textbf{0.3107}     &\textbf{0.4629}     & 1.752     \\\cline{2-6}
         &  GAN-        &0.0138     &0.4357     &0.4915     & 2.3343     \\\cline{2-6}
         &  Random      &0.5847     &1.1393     &5.2605     & 29.363    \\\cline{2-6}
         &  Template    &0.0219     &0.5433     &0.6973     & \textbf{0}         \\\hline
         \multirow{4}*{AN-S}
         &  GAN         &\textbf{0.0169}     &\textbf{0.4721}     &\textbf{0.5498}     & 2.3374\\\cline{2-6}
         &  GAN-        &0.0296     &0.5232     &0.5979     & 3.3324     \\\cline{2-6}
         &  Random      &0.5877     &2.8232     &6.0603     & 27.9345     \\\cline{2-6}
         &  Template    &0.0497     &0.5467     &0.6799     & \textbf{0}         \\\hline
         \multirow{4}*{AN-SL}
         &  GAN         &\textbf{0.0291}     &\textbf{0.6326}     &0.6171     & 3.876     \\\cline{2-6}
         &  GAN-        &0.0321     &0.6445     &0.6312     & 4.942     \\\cline{2-6}
         &  Random      &0.6571     &2.4812     &4.0639     & 22.6485    \\\cline{2-6}
         &  Template    &0.0513     &0.7892      &\textbf{0.6136}    & \textbf{0}         \\\hline
         
    \end{tabular}
    \caption{Query Load Variance Evaluation}
    \label{tab:query_load_exp1}
\end{table}

\paragraph{Experiment~1}
Table~\ref{tab:query_load_exp1} shows the distribution differences in several features between the query load generated by different methods and the input query load.

The Sequences MDD column shows the character level similarity between the generated query load and the input query load. It can be seen that our method achieves the best results for all four query loads, followed by the template method, and the random method generates query loads that are more different from the input query loads. This is because our method learns the distribution of the input query load and generates new query load that is similar to it.
Note that the full model GAN is better than GAN- for every query load. This illustrates the additional information: the cost and cardinality of the query load improve the performance of the model. At the same time, as the complexity of the query load increases, in particular the length of the query increases, leading to an increase in the sequence MMD. This suggests that the more complex the query load, the more difficult it is to learn its distribution. Despite this, our model still kepdf the sequential MMD at a low level.

The cardinality and cost WD columns show the distance between the distribution of the generated query load and the input query load in terms of cardinality and cost. In most cases, our full model gives the best results, followed by the GAN-, template and random methods. Note that the query load generated by the template method achieves the smallest WD for cost distribution in the ANN-N query load, probably because for longer queries, changing only the constants in the query does not have a significant impact on query cost.

The Length WD column shows the distance between the generated query load and the input query load in terms of length distribution. Since the template method only changes the constants in the input query load and does not change the length, the Length WD of the template method is 0. Except for the template method, our full model achieves the best results on the four query loads, followed by GAN- and random methods.

In summary, the query loads generated by GAN is more similar to the input query loads because our method learns the distribution of the input query load better. Besides, template has smaller errors than random because templates have been extracted from the input query load which partially contain information about the distribution of input queries.

\paragraph{Experiment~2}

\begin{figure}[htp]
	\centering
	\subfigure[Aggregation Functions]{
		\includegraphics[width=0.48\textwidth]{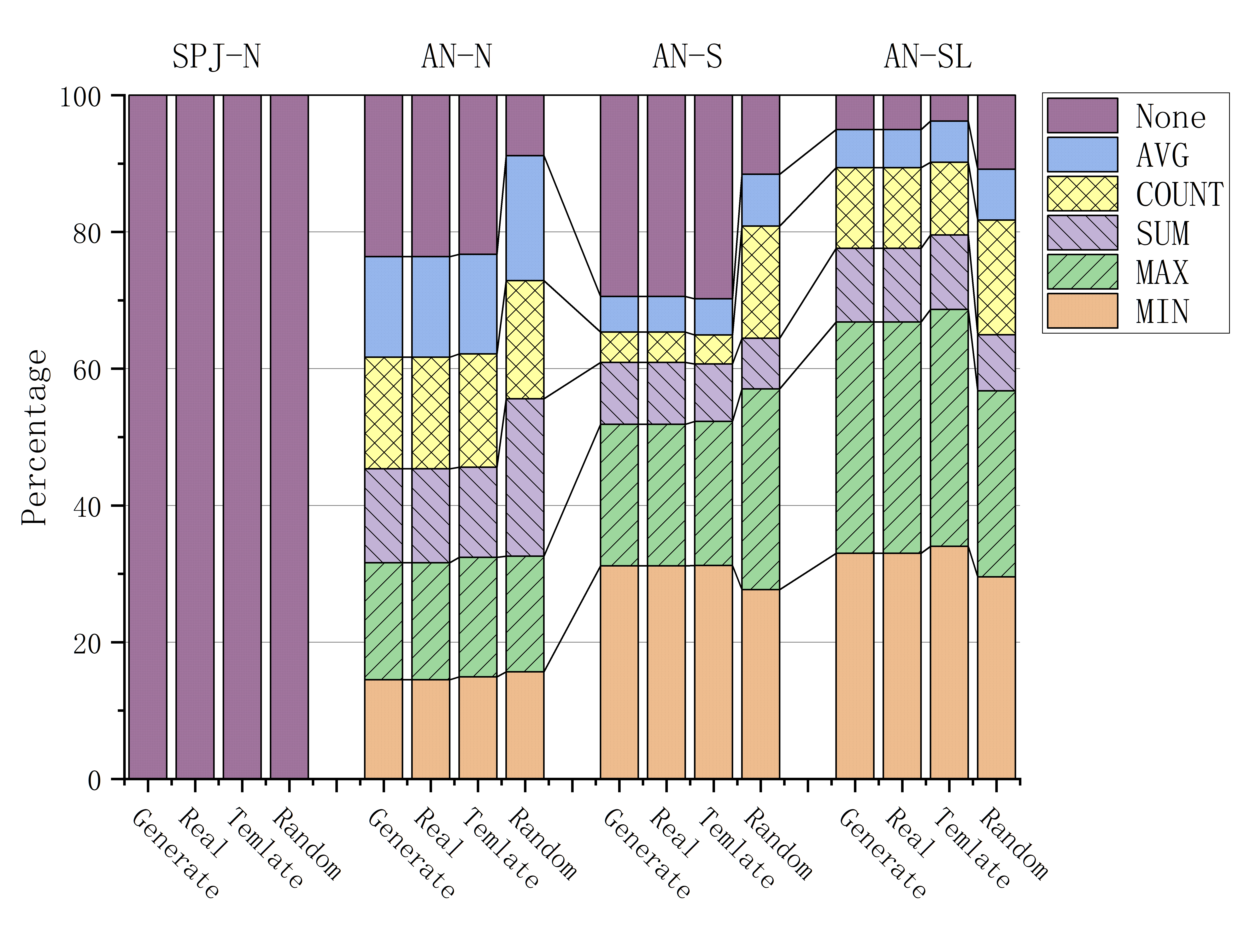}
	}%
	\subfigure[Operators]{
		\includegraphics[width=0.48\textwidth]{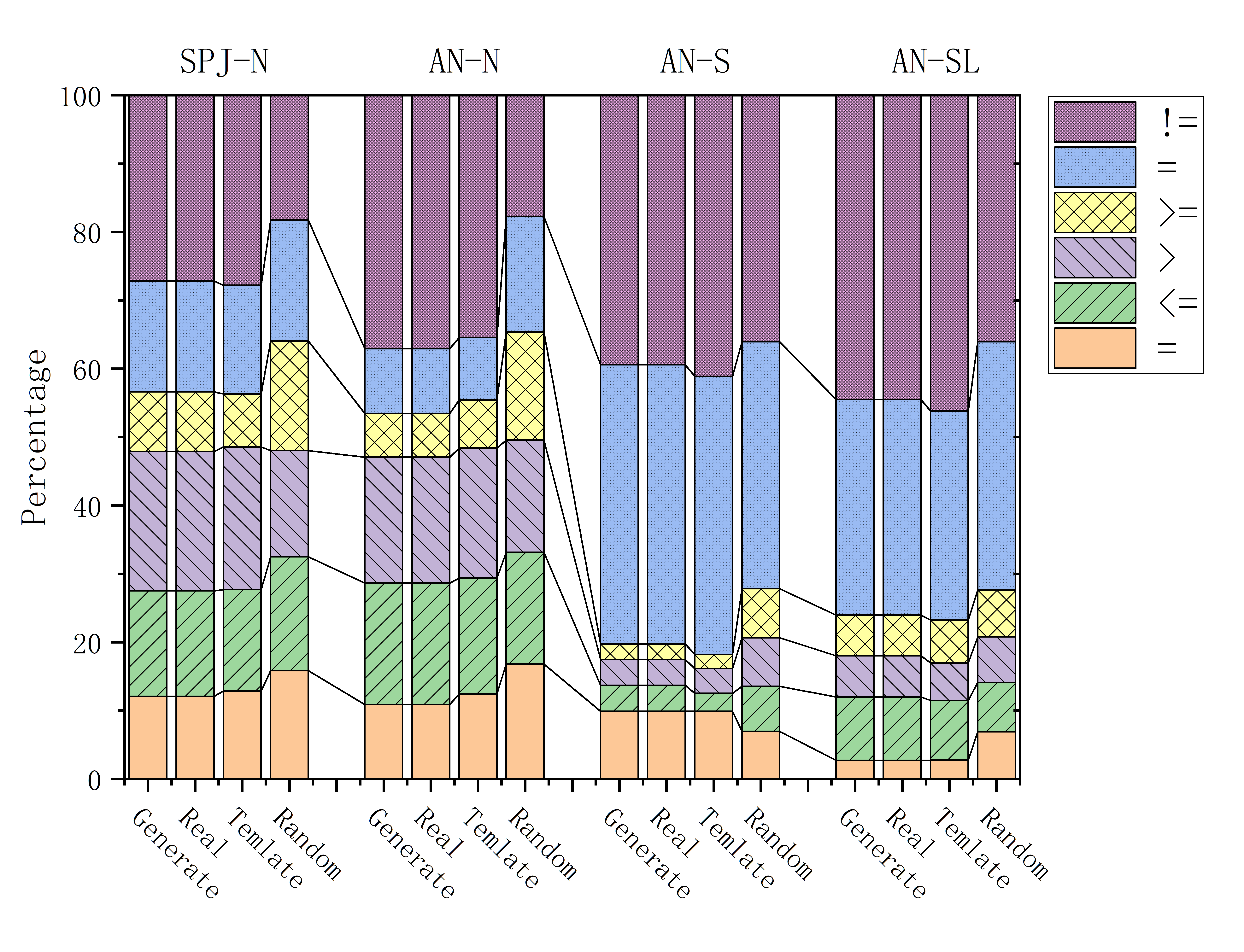}
	}
    \subfigure[Nested Queries]{
		\includegraphics[width=0.47\textwidth]{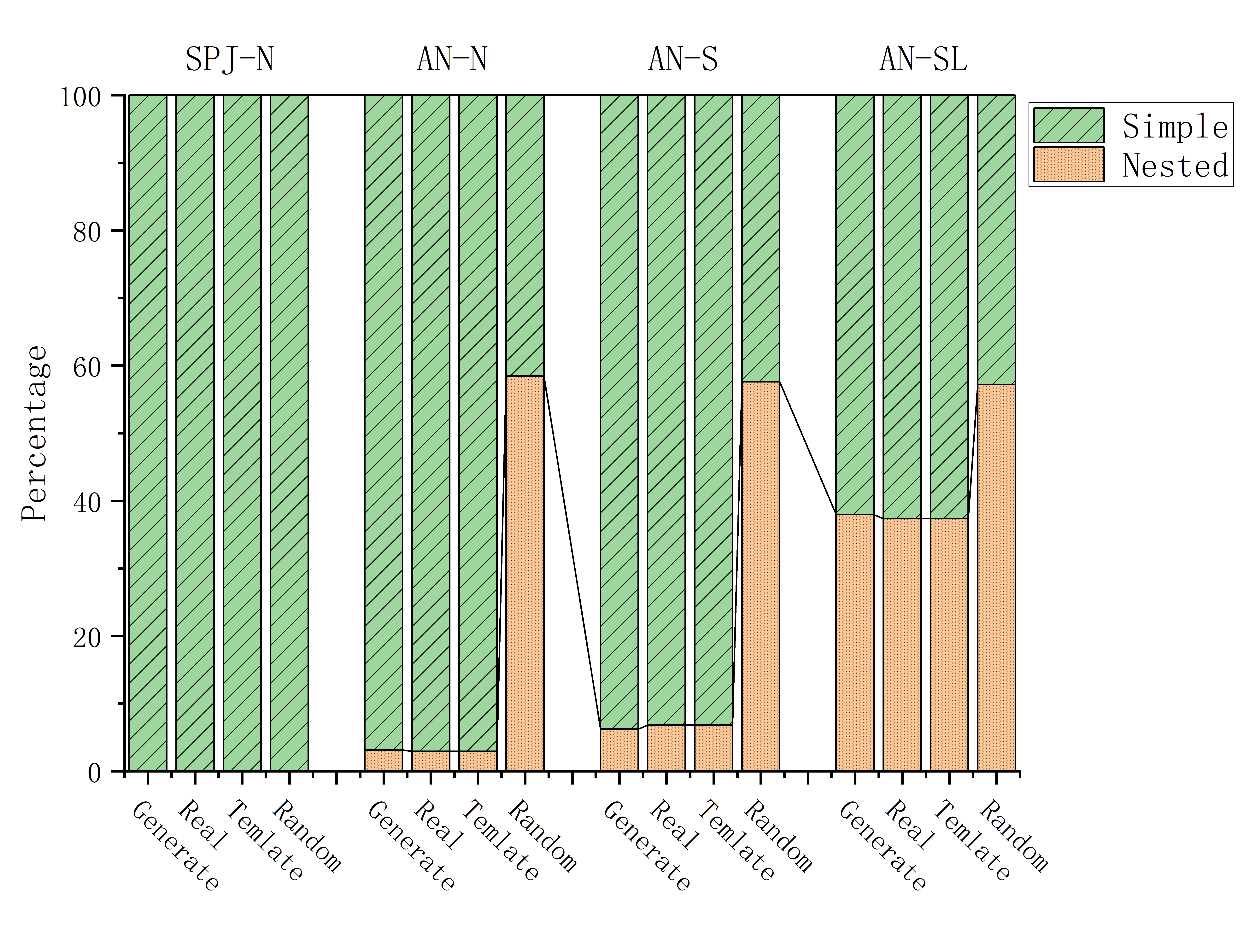}
	}
    \subfigure[Joined Tables]{
		\includegraphics[width=0.48\textwidth]{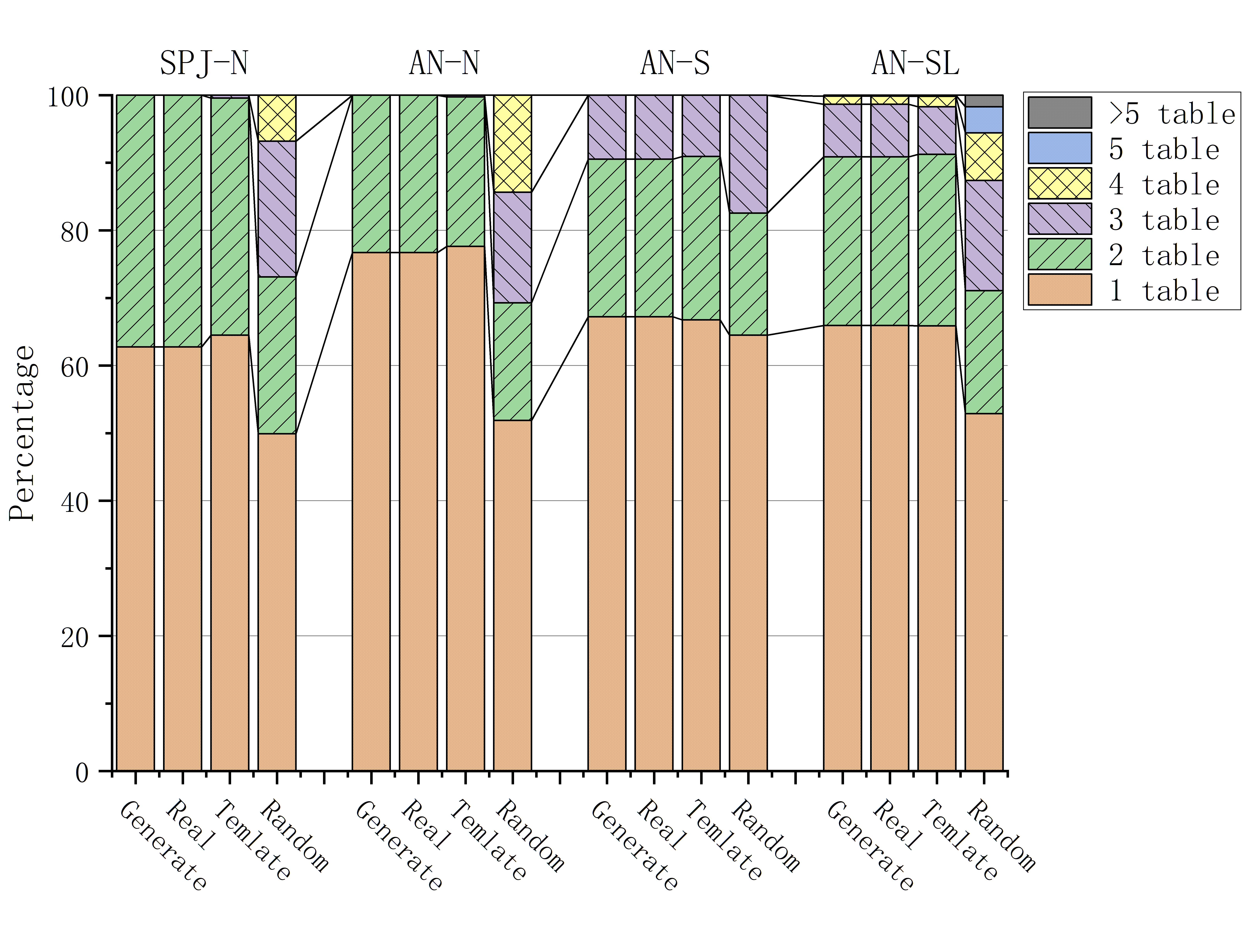}
	}
	\caption{Correlations between Occurrences of Attributes in Query Loads.}
	\label{fig:exp2}
\end{figure}

In this experiment, we consider the difference between the input query load and the generated query load for four features: aggregation functions, operators, nested queries, and number of joined tables. The result is shown in Fig~\ref{fig:exp2}. Each sub-figure is divided into four groups representing the results of the experiments across the four query loads. In each group, the first column (denoted as GAN) shows the distribution of query load generated using our method, the second column (denoted as Real) shows the input query load and the third column (denoted as Template) shows the query load generated using the template method, and the last column (denoted as Random) shows the query load generated using the random method.

Figure~\ref{fig:exp2}(a) shows the distribution of the aggregation functions across the four query loads. Since the template method does not change the query structure, the query load generated by the template method is the same as the input query load. Excluding the template method, our method has the most similar distribution to the input query load, and the worst is the random method. 

Note that in the AN-N group, the query load generated by the random method has essentially the same proportion of each aggregation function, which is because the random method has the same probability of choosing each aggregation function. However, the proportions of AVG and SUM aggregates are smaller in the AN-S and AN-LS group because AN-S and AN-LS introduce string data and AVG and SUM aggregation functions do not support strings. Our semantic module restricts the choice of AVG and SUM aggregation functions.

Figure~\ref{fig:exp2}(b) shows the distribution of the operators. The template query load is most similar to the input query load, followed by our method and finally the random method. Similar to the aggregate functions, after adding string data, we only support the "=" and "! =" operators for them, resulting in a skewed distribution of operators for the query load generated by the random method.

Figure~\ref{fig:exp2}(c) and (d) show the distribution of the nested queries and joined tables. Our method beats random method again.

\begin{figure}[htp]
	\centering
	\subfigure[Real]{
		\includegraphics[width=0.40\textwidth]{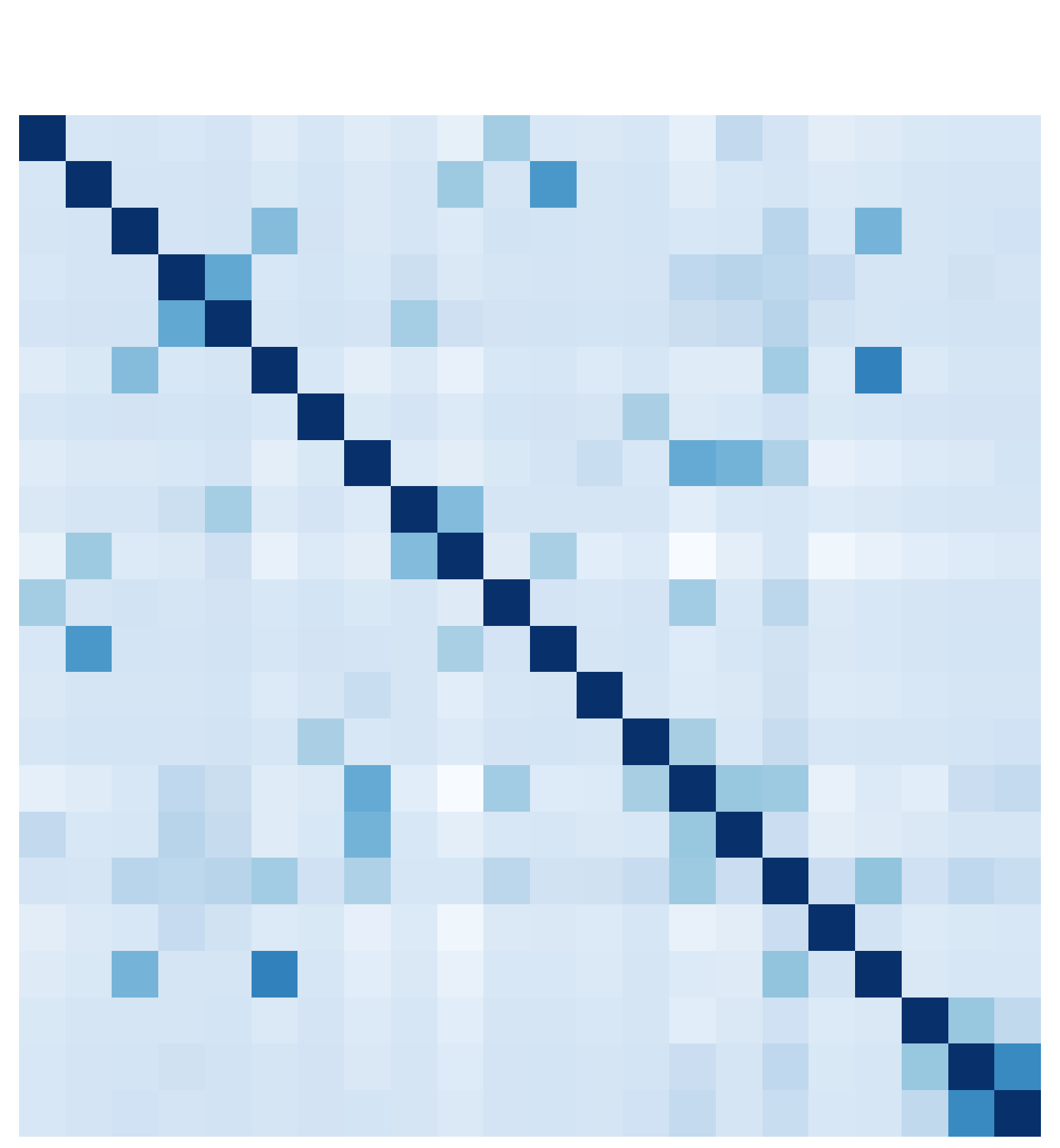}
	}%
	\subfigure[GAN]{
		\includegraphics[width=0.40\textwidth]{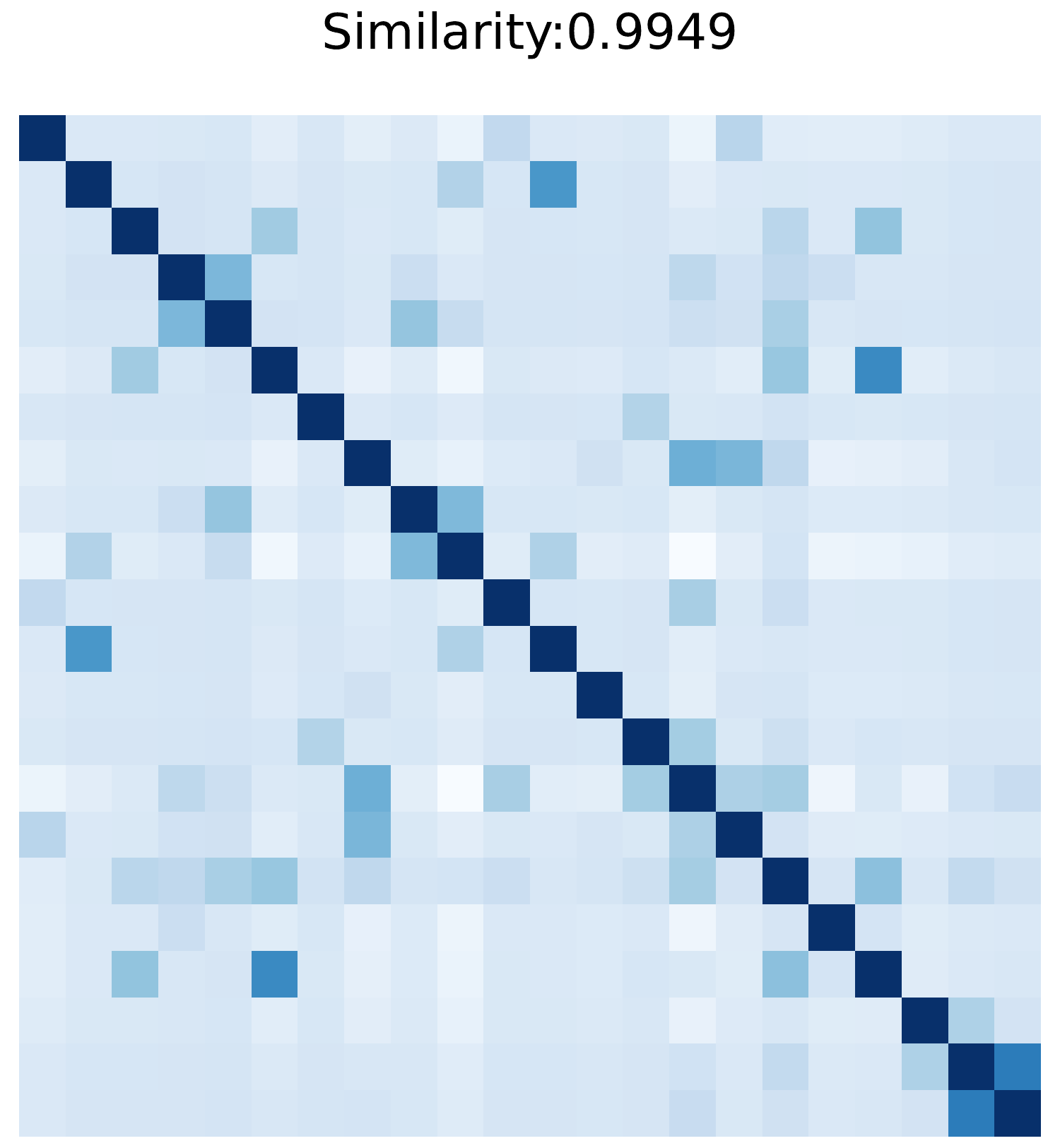}
	}
    \subfigure[Random]{
		\includegraphics[width=0.40\textwidth]{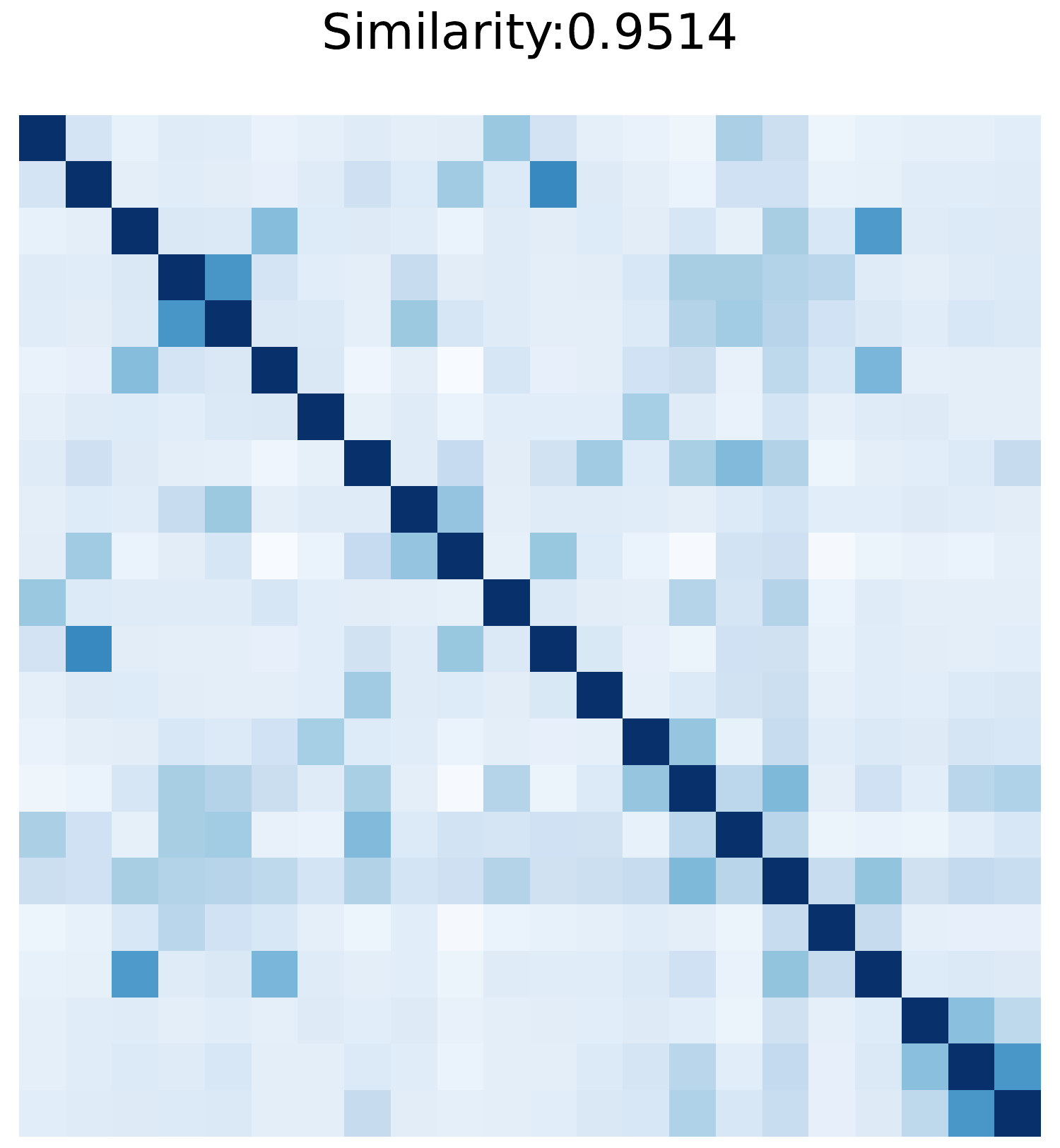}
	}
    \subfigure[Template]{
		\includegraphics[width=0.40\textwidth]{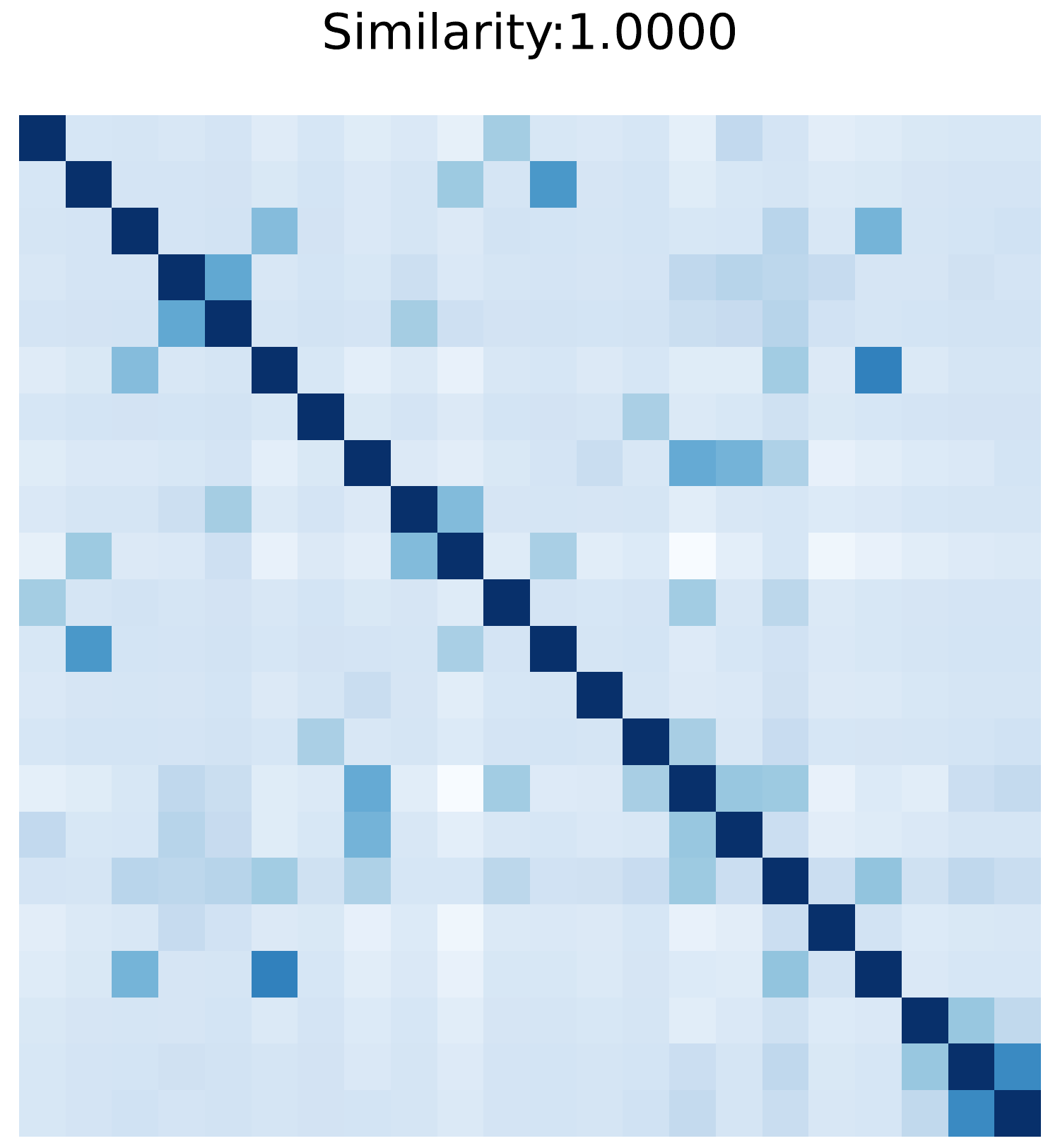}
	}
	\caption{HeatMap.}
	\label{fig:heatmap}
\end{figure}

\paragraph{Experiment~3}
In this experiment, we consider the correlation between the occurrences of two attributes in a query load as an important feature. Specifically, for an attribute $A$, let $o_i(A) = 1$ if $A$ appears in the $i$-th query of a query load; otherwise, $o_i(A) = 0$. Then, for two attributes $A$ and $A'$, we compute Pearson's correlation coefficient between $o_i(A)$ and $o_i(A')$. Fig.~\ref{fig:heatmap}(a) shows the correlation heatmap of the input query load and Fig.~\ref{fig:heatmap}(b) is the query load generated by our GAN method. It can be seen that two attribute correlation matrices are very similar. After concatenating all row vectors of two matrices, respectively, we can compute the cosine similarity of two concatenated vectors, which equals to 0.9949. It means that two correlation matrices are almost the same. Fig.~\ref{fig:heatmap}(c) shows the results obtained for the query load generated by random method and the similarity is 0.9514.

\subsection{Evaluation of Downstream Task}
One application of query generation is to expand the query load. In this experiment, we expand the query load using different methods and then apply the expanded query load to a downstream task: cardinality estimation. For cardinality estimation, we use the q-error metric, defined as:
$$
Qerror = \frac{1}{n}\sum_{i=1}^n \frac{\max{(y_i,\hat{y_i})}}{\min{(y_i,\hat{y_i})}}
$$
where $y_i$ denotes the ground-truth value and $\hat{y_i}$ is the estimated one.

We choose the query-driven cardinality estimator MSCN~\cite{kipf2018learned}, which use a multi-set convolutional network (MSCN) to estimate cardinality. We generate 3000 queries using a randomly initialised generator and enter them into the database for execution to obtain their real cardinality. 2000 queries are used as the training set (this training set is small, it needs to be expanded) and 1000 queries are used as the test set.

Next, we expand the training sets to 10000 queries by generating 8000 queries each based on the original training set using GAN, template and random methods. To compare with the generation methods, we also expanded the origin query load to 10,000 by replication. Four MSCNs were trained using the expand training sets with the same configuration, and tested using the same test set. The results are shown in Table~\ref{tab:qerror}.

\begin{table}[h]
    \centering
    \begin{tabular}{|c|c|c|c|c|c|c|}\hline
                &median     &mean           &90th       &95th       &99th       &max  \\ \hline
    Real        &1.307      &9.249          &6.239      &14.072     &92.041     &2265.9 \\ \hline
    GAN         &1.302      &4.269          &4.876      &9.596      &45.291     &687.1 \\\hline
    Template    &1.887      &33.145         &9.637      &21.817     &142.483    &22432.3 \\\hline
    Random      &3.848      &122.084        &111.826    &339.341    &2509.465   &16668.8 \\\hline
    \end{tabular}
    \caption{MSCN Qerror on Different Training set}
    \label{tab:qerror}
\end{table}

Overall, the best results were obtained with the model trained on the GAN expanded training set. For the mean q-error value, GAN method is 2.6 times more accurate than the model trained using only the real load; for the max q-error value, our method is 3.3 times more accurate. This is because our model can learn the distribution of the input query load and generate new query loads with a similar distribution. At the same time, the stochastic element of the GAN allows the GAN to generate new queries that are little different from the input query, which in turn allows for high quality extension of the data.

The q-error of MSCN using random and template methods to expand the training set is even higher than the case of using only the real query load. We suspect that the model uses a large number of queries with different distributions from the real query load as training data, causing the model to fit the wrong distribution and map the wrong queries to the cardinality. When tested, the model does not map the real queries to the correct cardinality.

\subsection{Evaluation of Our Method}
\label{sub:model-evaluation}

In this section, we evaluate the effectiveness of various components of our query generation method.

\begin{figure}[htp]
	\centering
	\subfigure[Update]{
		\label{fig:loss}
		\includegraphics[width=0.41\textwidth]{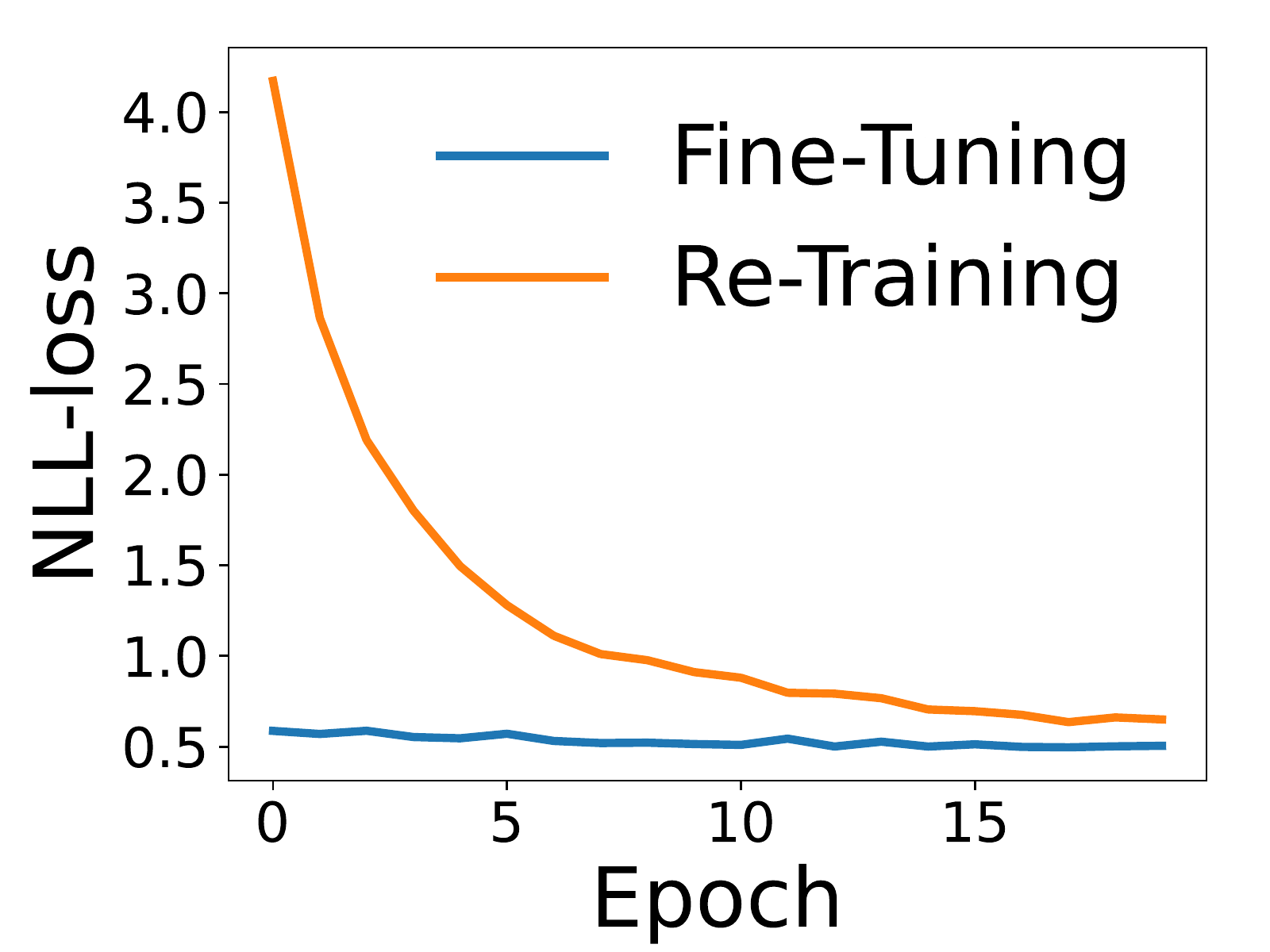}
	}%
	\subfigure[Syntax and Semantics]{
		\label{fig:rules}
		\includegraphics[width=0.41\textwidth]{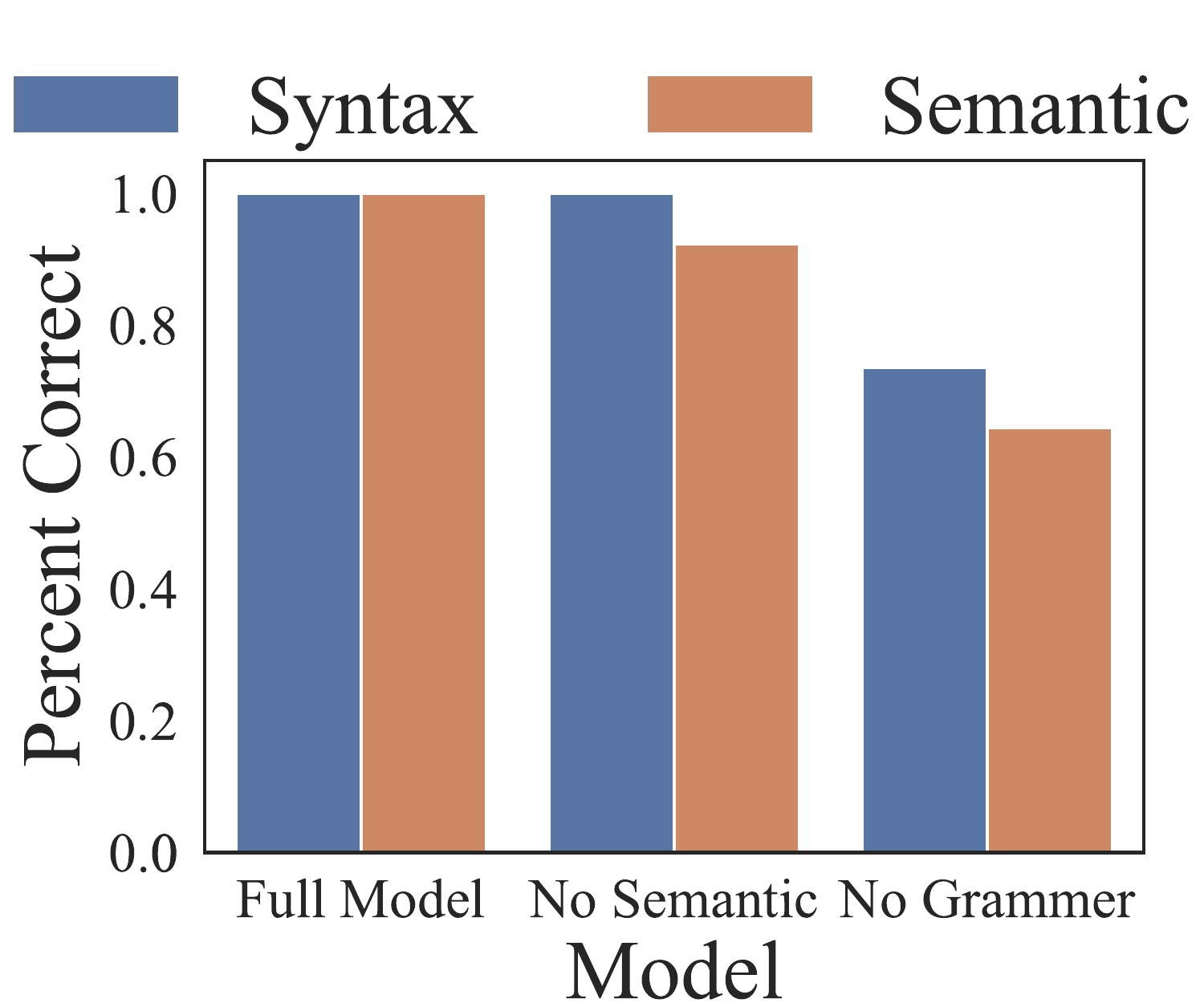}
	}
	\caption{Evaluation of Our Method.}
	\label{fig:components}
\end{figure}

\paragraph{Experiment~5}
We evaluate the impacts of the syntactic and semantic modules in GAN. Fig.~\ref{fig:rules} shows the fraction of generated queries that are syntactically correct or semantically correct (or both). The queries generated by our GAN model are all correct in syntax and semantics. When the syntax module is disabled, all the generated queries are syntactically correct, but only 92.3\% is semantically correct. When both the syntax module and the semantic modules are disabled, only 64.4\% of the generated queries are correct in semantic.
\paragraph{Experiment~6}
We also tested the model's ability to update when the query load changed. We adopt two updating strategies, retraining and fine-tuning. Fine-tuning is to continue training with new query loads based on the current model. To compare the differences between two update strategies, we synthesize two query loads A and B consisting of 2000 queries in the query length range of [0,40] and [40,100]. The results are shown in Fig.~\ref{fig:loss}. We first train a GAN model using load A, and then adopt two strategies to update for load B. The fine-tuning strategy stabilizes much faster than the retraining. Fine-tuning reaches a low nll-loss in 3 epochs, whereas retraining takes 15 epochs. This may be because the pre-trained model retains some basic information about the query. When the query load slightly changes, fine-tuning can help the model quickly learn the new query load distribution.

\section{Related Work}
\label{sec:related-work}

Existing query generation methods can be broadly divided into three categories: random generation, template-based generation, and reinforcement-learning-based generation.

\paragraph{Random Methods} Query loads generated by random methods are mainly used for large-scale random error diagnosis and testing of databases. Typical methods are RAGS~\cite{slutz1998massive} and SQL-SMITH~\cite{sqlsmith}. RAGS generates SQL statements by walking a stochastic parse tree and printing it out. SQL-SMITH generates a random query abstract syntax tree(AST) and then convert it into an SQL query by the formatting tool. While the queries generated by random methods conform to SQL syntax, they are completely random. Even SQL-SMITH may generate semantically incorrect queries, causing the query to fail to execute, or result in an empty execution.

\paragraph{Template-based Methods} Template-based methods use some given query templates and adjust the values in the predicates to generate queries that satisfies certain constraints. Bruno et al.~\cite{bruno2006generating} used a heuristic hill-climbing algorithm to search in the predicate parameter space to generate queries satisfying the cardinality constraint. Chaitanya et al.~\cite{mishra2008generating} divided the $d$-dimensional predicate space into $k^d$ smaller units on average, performed a unit-level search in the $d$-dimensional space, and selected the $k$ units with the highest scores to re-divide and search. Even though template-based methods can generate queries that satisfy certain constraints, they still cannot generate new queries that have a similar distribution to a given query load.

\paragraph{Reinforcement-learning-based Methods}
Query generation methods based on reinforcement learning (RL) can generate queries that satisfy certain constraints without given queries. A typical method is LearnedSQLGen~\cite{zhang2022learnedsqlgen}. LearnedSQLGen uses the actor-critic reinforcement learning network structure and generates queries by choosing appropriate token at each time step.  Furthermore, to guarantee the syntactic and semantic correctness of the generated queries, LearnedSQLGen uses a finite state machine to limit the action space. However, LearnedSQLGen does not capture the distribution of a given query load.

Another line of related work is text sequence generation, a widely studied technology in natural language processing (NLP). Query generation is essentially a text sequence generation problem because a query is expressed as an SQL statement. Long short-term memory networks (LSTM)~\cite{hochreiter1997long} and recurrent neural networks (RNN) have shown excellent performance in text generation. The most common method for training RNNs is the maximum likelihood method~\cite{salakhutdinov2015learning}. However, Bengio et al.~\cite{bengio2015scheduled} pointed out that the maximum likelihood method can produce exposure bias in the inference stage. SeqGAN~\cite{yu2017seqgan} introduces the GAN structure into text generation and uses the Monte Carlo method of reinforcement learning to train the generator, which solves the problem of exposure bias.

Many sequence generatio`n methods are fully stochastic processes~\cite{hochreiter1997long,yu2017seqgan} that select appropriate tokens from a token space step by step. Queries generated in this way may have syntax errors. For example, we are generating an SQL query and the generator has generated \texttt{"SELECT *"}. In the next step, the generator selects \texttt{*} in the token space, which results in an incorrect query \texttt{"SELECT **"}. Even if we guarantee that the query is syntactically correct, semantic errors may occur. In SQL, common semantic errors are mismatched operand types, meaningless predicate expressions, and inappropriate table joins. Semantic errors can lead to empty query results, unwanted query results, or even queries that fail to execute. TreeGAN~\cite{liu2018treegan} adds context-free grammar to the sequence generation process, which can generate grammar-compliant text. The deep-learning-based text generation methods provide the technical basis for studying our query generation method.

\section{Conclusion}
\label{sec:conclusion}

We propose a query generation method based on generative adversarial networks (GAN). The GAN model well captures the distributions of various features of the input queries. In addition, context-free grammar and semantic rules are adopted to guide query generation so that the generated queries are guaranteed to be correct in both syntax and semantics. The experiments verify that our method can generate new query loads that are very similar to the input query load.

\bibliographystyle{splncs04}
\bibliography{ref}

\end{document}